\documentclass[runningheads]{llncs}
\usepackage[T1]{fontenc}
\usepackage{graphicx}
\usepackage{hyperref}
\usepackage{color}

\let\lncsproof\proof \let\lncsendproof\endproof \let\lncsqed\qed
\let\proof\relax\let\endproof\relax

\usepackage{amsmath,amsthm}

\let\proof\lncsproof \let\endproof\lncsendproof \let\qed\lncsqed

\usepackage{mathtools}

\usepackage{amsfonts}

\usepackage[inline]{enumitem}
\usepackage{xspace}
\usepackage{nicefrac}
\usepackage{enumitem}

\usepackage{xcolor}

\usepackage{colortbl}

\usepackage{tabularx}
\usepackage{multirow}
\usepackage{booktabs}
\usepackage{array}

\usepackage{tcolorbox}
\usepackage{soul}

\usepackage{cleveref} 

\newtheorem{observation}[]{Observation}

\newtheorem{clm}{Claim}

\newcommand{\sse}{\subseteq}

\newcommand{\sm}{\setminus}

\usepackage{mathrsfs}

\newcommand{\cU}{\ensuremath{\mathcal{U}}}
\newcommand{\C}[1]{\ensuremath{\mathcal{#1}}}
\newcommand{\Co}[1]{\ensuremath{\mathcal{#1}}}

\newcommand{\X}[1]{\ensuremath{\mathscr{#1}}}
\newcommand{\B}[1]{\ensuremath{\mathbb{#1}}}

\newcommand{\cF}{\ensuremath{\mathcal{F}}}
\newcommand{\Oh}{\mathcal{O}\xspace}

\newcommand{\voters}{\ensuremath{\mathscr{V}}\xspace}
\newcommand{\candidates}{\ensuremath{\mathcal{C}}\xspace}

\newcommand{\vp}{\ensuremath{{\textrm R}}\xspace}
\newcommand{\cp}{\ensuremath{{\textrm B}}\xspace}

\newcommand{\OurCC}{{\sc Obnoxious-Egal-CC}\xspace}

\newcommand{\defproblem}[3]{
  \vspace{1mm}
\noindent\fbox{
  \begin{minipage}{\textwidth}
  \begin{tabular*}{\textwidth}{@{\extracolsep{\fill}}lr} #1 \\ \end{tabular*}
  {\bf{Input:}} #2  \\
  {\bf{Question:}} #3
  \end{minipage}
  }
  \vspace{1mm}
}

\renewcommand{\problem}{{\sc Obnoxious Egalitarian Median Committee Selection}\xspace}

\newcommand{\problemshort}{{\sc Obnox-Egal-Median-CS}\xspace}

\newcommand{\nph}{{\sf \textup{NP}}-hard\xspace}

\newcommand{\fpt}{\textsf{\textup{FPT}}\xspace}
\newcommand{\fptas}{\textsf{\textup{FPT}-\textup{AS}}\xspace}

\usepackage{tikz}
\usepackage{xkeyval}
\usepackage[textsize=small]{todonotes}

\newcommand{\ma}[1]{\todo[color={pink!40}]{#1} }

\usepackage{comment}
\excludecomment{suppress}

\newcommand{\hide}[1]{}

\newcommand{\Ma}[1]{\textcolor{magenta}{#1}}

\newcommand{\etal}{et al.}
\newcommand{\yes}{{\rm yes}}

\newcommand{\cP}{\mathcal{P}}
\newcommand{\arc}{\mathsf{arc}}
\newcommand{\cA}{\mathcal{A}}
\newcommand{\seg}[1]{\overline{#1}}
\newcommand{\btrue}{\texttt{true}\xspace}
\newcommand{\bfalse}{\texttt{false}\xspace}
\newcommand{\Real}{\mathbb{R}}

\begin{document}
\title{When far is better: The Chamberlin-Courant approach to obnoxious committee selection}
%
%
\author{Sushmita Gupta\inst{1} \and
Tanmay Inamdar \inst{2} \and
Pallavi Jain\inst{2} \and Daniel Lokshtanov\inst{3} \and Fahad Panolan\inst{4} \and Saket Saurabh\inst{1,5}}
\authorrunning{S. Gupta et al.}
%
\institute{The Institute of Mathematical Sciences, HBNI, Chennai \\
\email{\{sushmitagupta,saket\}@imsc.res.in} \and
Indian Institute of Technology Jodhpur \\
\email{\{taninamdar,pallavi\}@iitj.ac.in} \and
UC Santa Barbara\\
\email{daniello@ucsb.edu} \and
School of Computing, University of Leeds \\
\email{f.panolan@leeds.ac.uk}
\and
University of Bergen}
\maketitle              
\begin{abstract}
\hide{The abstract should briefly summarize the contents of the paper in
150--250 words.}

Classical work on metric space based committee selection problem interprets distance as “near is better”. In this work, motivated by real-life situations, we interpret distance as “far is better”. Formally stated, we initiate the study of ``obnoxious'' committee scoring rules when the voters' preferences are expressed via a metric space. To accomplish this, we propose a model where \emph{large distances imply high satisfaction} (in contrast to the classical setting where shorter distances imply high satisfaction) and study the egalitarian avatar of the well-known Chamberlin-Courant voting rule and some of its generalizations. For a given integer value $\lambda$ between $1$ and $k$, the committee size, a voter derives satisfaction from only the $\lambda$th favorite committee member; the goal is to maximize the satisfaction of the least satisfied voter. For the special case of $\lambda=1$, this yields the egalitarian Chamberlin-Courant rule. \hide{Gupta \etal [IJCAI 2021] proved that finding a solution, i.e $k$-sized winning committee using egalitarian median rule is {\sf NP}-hard for all $\lambda<k$.}  In this paper, we consider general metric space and the special case of a $d$-dimensional Euclidean space. 
%

We show that when $\lambda$ is $1$ and $k$, the problem is 
polynomial-time solvable in $\mathbb{R}^2$ and general metric space, respectively. However, for $\lambda=k-1$, it is \nph even in $\mathbb{R}^2$. Thus, we have ``double-dichotomy'' in $\mathbb{R}^2$ with respect to the value of $\lambda$, where the extreme cases are solvable in polynomial time but an intermediate case is \nph. 
Furthermore, this phenomenon appears to be ``tight'' for $\mathbb{R}^2$ because the problem is \nph for general metric space, even for $\lambda=1$. Consequently, we are motivated to explore the problem in the realm of (parameterized) approximation algorithms and obtain positive results. Interestingly, we note that this generalization of Chamberlin-Courant rules encodes practical constraints that are relevant to solutions for certain facility locations.

\end{abstract}
%


\newpage

\section{Introduction}


Initiated in the 18th century, the multiwinner election problem, also known as the committee selection problem, has been central to social choice theory for over a century~\cite{ChamberlinCourant,Thiele1895,Monroe1995} and
and in the last decade and a half it has been among the most well-studied problems in computational social choice~\cite{Aziz2019,mwchapter,brandt2016tournament,brams2007minimax}. In this problem,  given a set of candidates $\mathcal{C}$, a profile $\mathcal{P}$ of voters' preferences, and an integer $k$; the goal is to find a $k$-sized subset of candidates (called a {\it committee}) using a multiwinner voting rule. The committee selection problem has many applications beyond parliamentary elections, such as selecting movies to be shown on a plane, making various business decisions, choosing PC members for a conference, choosing locations for fire stations in a city, and so on. For more details on the committee selection problem, we refer the reader to~\cite{mwchapter,lackner2023multi}. 


The Chamberlin-Courant (CC) committee is a central solution concept in the world of committee selection. Named after Chamberlin and Courant~\cite{ChamberlinCourant}, it is derived from a\hide{{\it committee scoring rule}--where we associate a \emph{satisfaction} value to each voter towards a committee as a function that considers relative positions of the committee members in the preference of the voter} multiwinner voting rule where the voter's preference for a given $k$-sized committee is evaluated by adding the preference of each voter for its {\it representative}, the most preferred candidate in the committee.\hide{\footnote{Conversely, the subset of voters who are being represented by a certain committee member is called the {\it virtual district} of that candidate in that committee.}\todo{@Sushmita: why did we add this footnote? \Ma{to establish a correspondence between committee members and voters.We may skip}}} The CC committee is one with the highest value. There has been a significant amount of work in computational social choice centered around this concept  and has to date engendered several {\it CC-type rules} that can be viewed as a generalization of the above. \hide{of the notion of representation} Specifically, {\it ordered weighted average}(OWA) operator-based rules such as the {\it median scoring rule}, defined formally later, \hide{and the {\it best scoring rule},}  \cite{SkowronFL16,aziz2018egalitarian} can be seen as a direct generalization of CC. Moreover, there are other notions of generalization based on the preference aggregation principle: the original CC rule is {\it utilitarian}, that is, it takes the summation of each voter's preference value toward its representative,~ \cite{SkowronFL16,DBLP:conf/aaaiBredereckF0KN20,DBLP:conf/ijcai/Gupta00T21}. The {\it egalitarian} variant studied by Aziz~\etal\cite{aziz2018egalitarian} and Gupta~\etal\cite{DBLP:conf/ijcai/Gupta00T21} is one where only the least satisfied voter's preference value towards its representative is taken. Clearly, there could be many other variants where some other aggregation principle is considered. We refer to all these variants collectively as the CC-{\it type} rules and the egalitarian variants as the {\it egalitarian} CC-{\it type} rules. The egalitarian rules, also known as Rawlsian rules, are based on Rawls's theory of ``justice as fairness'' that favors equality in some sense by maximising the minimum satisfaction. Egalitarian rules are very well-studied in voting theory~\cite{klein2023egalitarian,aziz2018egalitarian,DBLP:conf/ijcai/Gupta00T21,DBLP:conf/ijcai/DeltlFB23,DBLP:conf/ijcai/SreedurgaBN22}. 



\hide{
\paragraph{{\bf Significance of Egalitarian Rules.}}
\Ma{The egalitarian rules, also known as Rawlsian rules, are based on Rawls's theory of ``justice as fairness'' that favors equality in some sense by maximising the minimum satisfaction. Egalitarian rules are well-studied in voting theory~\cite{klein2023egalitarian,aziz2018egalitarian,DBLP:conf/ijcai/Gupta00T21,DBLP:conf/ijcai/DeltlFB23,DBLP:conf/ijcai/SreedurgaBN22} and in recent times have been extended to committee selection,~\cite{aziz2018egalitarian,DBLP:conf/ijcai/Gupta00T21} } \todo{@Sushmita:do we need to add anything more?}\ma{this is enough!}}


In this paper, we consider the situation where the voters' preferences are expressed via a metric space, a natural setting in the facility location problem and spatial voting.
Furthermore, we consider egalitarian scoring rules, which aim to maximize the ``satisfaction'' of the least satisfied voter. Moreover, Gupta \etal~\cite{DBLP:conf/ijcai/Gupta00T21} study a wide range of egalitarian rules, called the {\it egalitarian median rule}, which is a generalization of the egalitarian CC rule\hide{({\sc Minimax CC-Multiwinner} in ~\cite{DBLP:journals/jair/BetzlerSU13})} and is defined as follows: for a voter $v$ and a committee $S$, let $pos_v^S$ (called a \emph{position vector}) be the vector of positions of candidates in $S$ in the ranking of $v$ in increasing order. For example, for the voter $v\colon a \succ b \succ d \succ c$ and set $S=\{c,d\}$, the vector $pos_v^S=[3,4]$. In the egalitarian median rule, given a value $1\leq \lambda \leq k$, the satisfaction of a voter $v$ for a committee $S$ is given by $m-pos_v^S[\lambda]$, where $m$ is the total number of candidates and $pos_v^S[\lambda]$ is the position value of the $\lambda^{th}$ candidate in $S$ according to $v$'s preference list. \hide{\ma{added this. do we need the next ?} That is, the satisfaction of a voter $v$ for a committee $S$ is, say, the Borda score of the $\lambda$th preferred candidate of $v$ in $S$ (any scoring function can be used instead of Borda score).} Note that when $\lambda=1$, we get the egalitarian Chamberlin-Courant scoring rule. Gupta \etal~\cite{DBLP:conf/ijcai/Gupta00T21} proved that the egalitarian median rule is \nph for every $\lambda<k$. 

In contrast to the aforementioned intractibility results for egalitarian CC-type rules, Betzler~\etal\cite{betzler2013computation} show that for $\lambda=1$, the egalitarian median rule is polynomial-time solvable for $1$-dimensional Euclidean preferences because such preferences are single-peaked. This motivates us to study egalitarian CC-type rules in the metric space setting that goes beyond dimension $1$. 

\paragraph{\bf{Preferences via a metric space.}} \hide{We focus our study on the case where the voters' preferences are encoded in a metric space.} For a given metric space, preferences are encoded in the following manner: each candidate and each voter is represented by a point in the metric space. In earlier works, distance is viewed as being inversely proportional to preference, that is, a voter is said to have a higher preference for candidates who are closer to her than those that are farther in the metric space,~\cite{mwchapter,SuriMultiwinner22,DBLP:journals/scw/ChenG21}. In this paper, we consider the opposite scenario where {\it distance is directly proportional to preference}, that is, a candidate who is \emph{farther away is more preferred} than the one who is closer. Inspired by {obnoxious facility location~\cite{ObnFL2,tamir1991obnoxious,church2022review}}, we call our problem {\it obnoxious committee selection}. Before we delve into the formal definition of our problem, we discuss the use of metric space to encode preference in earlier works. 



\begin{itemize}
\item The facility location problem is actually equivalent to the committee selection problem, where we assume that the closer facility is more preferred. The well-known {\sc $k$-center} problem (also known as the {\sc Minimax Facility Location} problem in metric space) is equivalent to the egalitarian CC committee selection problem when the voters' preferences are encoded in a metric space, where higher preference is given to the candidate that is closer. For some applications, it is natural to demand more than one facility in the vicinity, e.g., convenience stores, pharmacies, healthcare facilities, playgrounds, etc. This is known as the \emph{fault tolerant {\sc $k$-center}} problem and is captured by egalitarian median rules in a metric space~\cite{DBLP:journals/tcs/ChechikP15}. Facility location is among the most widely studied topics in algorithms, and we point the reader to some recent surveys~\cite{an2017recent,DBLP:conf/soda/Cohen-Addad0LS23,DBLP:conf/soda/GowdaPST23} on the topic and to \cite{DBLP:conf/ijcai/ChanFLLW21} for a survey on facility location in mechanism design.

\item In the spatial theory of voting, voters and candidates are embedded in the $d$-dimensional Euclidean space, and each voter ranks the candidates according to their distance from them~\cite{DBLP:journals/corr/abs-2302-08929,laslier2006spatial}. 

\hide{
\item Finally, we discuss a scenario which closely resembles the situation we want to study here. Consider the challenge of selecting a set of reviewers, let us call them the program committee, for a conference. Ideally, there should be at least three reviewers per submission who have no conflicts of interest. The co-author graph captures collaborative relations between various researchers. We can enrich this graph by encoding other types of conflicts as well. Moreover, the conflicts themselves may vary in terms of strong, moderate, and weak. Such information can be encoded as distances in a suitably defined metric space, where points represent submissions and reviewers and the distance between such points capturing the highest level of conflict between the authors of the submission and the reviewer with short distance implying high conflict. For eg, if a reviewer $r$ has a strong conflict with any of the authors of a submission $p$, then $d(p, r)=1$; if $r$ has only moderate to weak conflicts with the authors of $p$, then $d(p,r)=2$; if $r$ has only weak conflicts with the authors of $p$, then $d(p,r)=3$; and $d(p,r)\geq 4$, otherwise. Thus, the aforementioned challenge can be seen as finding a committee of size $k$ such that for each submission we want at least three reviewers that are at distance at least 4 from it. 
}

\end{itemize}

In the last few years, a fair amount of research centered on the theme of voting, committee selection, especially the CC rule, in metric spaces have appeared in theory and economics and computation venues~\cite{SuriMultiwinner22,DBLP:conf/sigecom/MunagalaSW21,MetricDistortion24,LowDistortionCS24,DBLP:journals/corr/abs-2302-08929,MunagalaSWW22,MunagalaMavrovEC23,AnariCR23}. Motivated by the applications stated above, we consider the general metric spaces as well as the Euclidean space for our study. Next, we discuss our motivation for studying \emph{obnoxious committee} selection before presenting the formal definition. 



\paragraph{\bf{Why Obnoxious Committee?}} The committee selection problem has been studied for the metric space in literature~\cite{DBLP:conf/aaai/ElkindFLSST17,DBLP:conf/aaai/GodziszewskiB0F21,DBLP:journals/tcs/KhullerPS00,DBLP:conf/aaai/SkowronE17,SuriMultiwinner22}. All these papers use the ``closer is better" perspective and thus candidates that are closer are preferred over those that are farther. Motivated by real-life scenarios where {\it every kind of facility is not desirable in the vicinity} such as is the case with factories, garbage dumps and so on, we want to study a problem which allows us to {\it restrict the number of facilities in the vicinity}. This is particularly relevant for facilities that bring some utility but too many leads to loss in value or even to negative utility. In order to design an appropriate solution concept for scenarios such as these we associate higher preference to facilities (i.e candidates) that are far and set the value of $\lambda$ in a situation-specific way. For example, if the local government is searching locations to build $k$ factories, then a solution with $\lambda=k$ where the various neighborhoods are voters and the potential locations are candidates should ensure that each of the $k$ factories is located far from every neighborhood. Moreover, for facilities such as garbage recycling , we can set $\lambda = k- \Co{O}(1)$ so that all but few facilities are located far from any neighborhood. Since the value of $\lambda$ can depend on $k$ (which is part of an input), we take $\lambda$ to be part of the input. Overall, we observe that as far as satisfaction is concerned, different facilities bring different levels of satisfaction depending on how many of them are in the vicinity. Consequently, it is desirable to have a model which is robust enough to capture this nuance. This translates to $\lambda$ being user defined, and is thus specified as part of the input to the problem.

\paragraph{{\bf Formal definition.}} We introduce some notation before giving a formal definition of the problem studied in this paper.
For a given metric space $\X{M}=(X, d)$, a point $x\in X$, a subset $S \sse X$, and $\lambda \in [k]$, we define $d^{\lambda}(x, S)$ to be the distance of $x$ to the $\lambda^{th}$ farthest point in $S$. To define this notion formally, we may sort the distances of a point $x$ to each $s \in S$ in non-increasing order (breaking ties arbitrarily, if needed), and let these distances be $d(x, s_1) \ge d(x, s_2) \ge \ldots \ge d(x, s_k)$. Then, $s_{\lambda} \in S$ is said to be the $\lambda^{\text{th}}$ farthest point from $x$ in $S$, and $d^{\lambda}(x, S) = d(x, s_{\lambda})$. Note that $d^1(x, S)$ is the distance of $x$ from a farthest point in $S$. For a point $p \in X$ and a non-negative real $r$, $B(p, r) \coloneqq \left\{q \in X: d(p, q) \le r \right\}$ denotes the ball of radius $r$ centered at $p$.


\defproblem{\problem~\\(\problemshort, in short)}{A metric space \X{M} consisting of a  set of voters, \voters, a set of candidates, \candidates; positive integers $k$ and $\lambda \in [k]$; and a positive real $t$.}{Does there exist a subset $S\sse \candidates$ such that $|S| = k$ and for each $v\in \voters$, $d^{\lambda}(v, S) \geq t$?}


When $\lambda=1$, we give the problem a special name, \OurCC, due to its similarity with the egalitarian CC rule. (Note that the egalitarian (resp. utilitarian) CC rule itself is the special case of the egalitarian (resp. utilitarian) median rule when $\lambda=1$.)




\paragraph{\bf Our Contributions.} In the following, we discuss the highlights of our work in this paper and the underlying ideas used to obtain the result. 


\begin{itemize}
\item \begin{sloppypar} We begin with studying \OurCC, that is, \problemshort with $\lambda=1$, and show that it is polynomial-time solvable when voters and candidates are embedded in $\mathbb{R}^2$ with Euclidean distances,\Cref{thm:polytime-plane}. To design this algorithm, we first observe that the above setting can be equivalently reformulated as the following geometric problem. Given $\voters$, and a set of equal-sized disks $\mathcal{D}$, find a $k$-size subset $\mathcal{D}' \subseteq \mathcal{D}$ such that no point of $\voters$ belongs to the common intersection region of $\mathcal{D}'$. Following that we use geometric properties of equal-sized disks to design an algorithm that uses dynamic programming to inductively build such a region. This algorithmic result contrasts with the intractability of the non-obnoxious version (the {\sc $k$-Center} problem) which is known to be \nph in $\mathbb{R}^2$. 
\item In \Cref{thm:apx-hardness-cc}, we consider \OurCC in general metric spaces. We show that it is \nph, and in fact,  the optimization variant is also hard to approximate beyond a factor of $1/3$ in  $f(k)\cdot n^{\Oh(1)}$ time, where $f$ is any computable function and $k$ is the committee size; and also W$[2]$-hard when parameterized by $k$. The latter implies that no algorithm with running time $f(k)n^{\Oh(1)}$ is likely to exist.
\end{sloppypar}
\item Notwithstanding these negative results, we show that \OurCC admit a factor 1/4 approximation algorithm that runs in polynomial time, \Cref{thm:polytime-approx}. In this algorithm, we first compute a ``$t/2$-net'' $S \subseteq \candidates$, i.e., $S$ satisfies the following two properties: (1) $d(c, c') > t/2$ for any distinct $c, c' \in S$, and (2) for any $c \not\in S$, there exists some $c' \in \X{M}$ such that $d(c, c') \le t/2$. Now, consider a point $p \in \voters$ and a $c^* \in \candidates$, such that $d(p, c^*) \ge t$. Then, by using the two properties of $S$, we argue that there exists a point in $c' \in S$ that is ``near'' $c$, and hence, ``far from'' $p$. More specifically, we can show that $d(p, c') \ge t/4$, leading to a $1/4$-approximate solution. 

\item Our work on \problemshort for $\lambda>1$ reveal that for $\lambda=k$, the problem can be solved in the polynomial time due to the fact that every committee member needs to be at least $t$ distance away from every voter. So, if possible, we can choose any $k$ candidates that are $t$-distance away from every voter; otherwise, a solution does not exist. The algorithm is same as the one in Proposition 3  in~\cite{aziz2018egalitarian}, but here we can have ties.   
%
 We show that for $\lambda = k-1$, \problemshort is \nph (\Cref{thm:r2-hardness}) even when the voters and candidates are points in $\mathbb{R}^2$. Furthermore, we show that the intractability results we have for \OurCC in \Cref{thm:apx-hardness-cc} carry forward to $\lambda>1$, as shown in Theorem~\ref{thm:apx-hardness}.


\item For an arbitrary value of $\lambda$ in $\Real^d$ space, we exhibit a \textit{fixed-parameter tractable approximation scheme}, that is, an algorithm that returns a solution of size $k$, in time \fpt in $(\epsilon, \lambda, d)$, such that for every point $v\in \voters$ there are at least $\lambda$ points in the solution that are at distance at least $(1-\epsilon)t$ from $v$, \Cref{thm:fpt-as}. Note that $\lambda\leq k$, thus, this algorithm is also \fpt in $(\epsilon, k, d)$. To obtain this result, we first observe that it is possible to refine the
idea of $t/2$-net further, and define a set of ``representatives'', if the points belong to a Euclidean space. In this setting, for any $0 < \epsilon < 1$, we can compute a candidate set $\mathcal{R}$ of representatives, such that for every relevant $c \in \candidates$, there exists a $c' \in \mathcal{R}$ such that $d(c, c') \ge \epsilon/2$. Moreover, $\mathcal{R}$ is bounded by a function of $\lambda, d$, and $\epsilon$. Thus, we can find an $(1-\epsilon)$-approximation by enumerating all size-$k$ subsets of $\mathcal{R}$, leading to the following result.

\end{itemize}


\paragraph{\bf{Related works.}}



Much of the research on multiwinner voting is concentrated on the computational complexity of computing winners under various rules, because for many applications it is crucial to be able to efficiently compute exact winners. As might be expected, computing winners under some committee scoring rules can be done in polynomial time (e.g., $k$-Borda~\cite{mwchapter}), while for many of the others the decision problem is \nph.

\begin{sloppypar}
Sufficient effort has gone towards applying the framework of parameterized complexity to these  problems. Some of the commonly studied parameters are the committee size $k$ and the number of voters, $n$. Indeed, this line of research has proven to be rather successful (see, e.g., \cite{bredereck2017robustness,DBLP:conf/aaaiBredereckF0KN20,faliszewski2017multiwinner,faliszewski2019committee,faliszewski2018multiwinner,aziz2018egalitarian,betzler2013computation,betzler2012studies,faliszewski2017bribery,yang2018parameterized,zhou2019parameterized,liu2016parameterized,aziz2014computational,misra2015parameterized,DBLP:conf/ijcai/Gupta00T21,DBLP:conf/ijcai/000122}). The problem has also been studied through the perspectives of approximation algorithms~\cite{DBLP:conf/sigecom/MunagalaSW21,DBLP:conf/atal/BrillFST19} and parameterized approximation algorithms~\cite{DBLP:journals/iandc/Skowron17,DBLP:journals/jair/SkowronF17,DBLP:conf/aaaiBredereckF0KN20}.
\end{sloppypar}


It is worth noting the similarities between our model and that of the \emph{fault tolerant} versions of clustering problems, such as $k$-\textsc{Center} or $k$-\textsc{Median} \cite{KumarR13,inamdar2020fault,chakrabarty2023fault}, also \cite{chakrabarty2017interpolating}. In the latter setting, the clustering objective incorporates the distance of a point to its $\lambda^{\text{th}}$ closest chosen center. Here, $\lambda \ge 1$ is typically assumed to be a small constant. Thus, even if $\lambda -1$ centers chosen in the solution undergo failure, and if they all happen to be nearby a certain point $p$, we still have some (upper) bound on the distance of $p$ to its now-closest center. Note that this motivation of fault tolerance translates naturally into our setting, where we want some (lower) bound on the distance of a voter to its $\lambda^{\text{th}}$ \emph{farthest} candidate, which may be useful of the $\lambda - 1$ farthest \emph{candidates} are unable to perform their duties.

\paragraph{\bf{Preliminaries}}


\hide{
For a point $x\in X$, a subset $S \sse X$ and $\lambda \in [k]$, we define $d^{\lambda}(x, S)$ to be the distance of $x$ to the $\lambda^{th}$ farthest point in $S$ (using a fixed tie-breaking rule if needed). Thus, $d^1(x, S)$ is the distance of $x$ from the farthest point in $S$. 



For the purpose of exposition, we identify the voters as {\it blue} points and the candidates as {\it red} points.

\defproblem{\problem~(\problemshort, in short)}{A metric space \X{M} consisting of a  set of voters, \voters, and a set of candidates, \candidates; integers $k$ and $\lambda \in [k]$; and a positive real $t$}{Does there exist a subset $S\sse \candidates$ such that $|S| = k$ and for each $v\in \voters$, $d^{\lambda}(v, S) \geq t$?}
}


	In the optimization variant of \problemshort, the input consists of $(\X{M}, \voters, \candidates, k, \lambda)$ as defined above, and the goal is to find the largest $t^*$ for which the resulting instance is a yes-instance of \problemshort, and we call such a $t^*$ the optimal value of the instance. We say that an algorithm has an approximation guarantee of $\alpha \le 1$, if for any input $(\X{M}, \voters, \candidates, k, \lambda)$, the algorithm finds a subset $S \subseteq \candidates$ of size $k$ such that for each $v \in \voters$, $d^{\lambda}(v, S) \ge \alpha \cdot t^*$. 




\paragraph{Brief primer on parameterized algorithms.} In parameterized algorithms, given an instance $I$ of a problem $\Pi$ and an integer $k$ (also known as the \emph{parameter}), the goal is to design an algorithm that runs in $f(k)\cdot |I|^{\Oh(1)}$ time, where $f$ is an
arbitrary computable function depending only on the value of $k$. Such an algorithm is said to be \emph{fixed-parameter tractable} (\fpt) with respect to the parameter $k$. Moreover, we will refer to such an algorithm as an \fpt algorithm. On the other hand, parameterized problems that are hard for the complexity class {\sf W[r]} for any $r\geq 1$ do not admit fixed-parameter algorithms with respect to that parameter, under the standard complexity assumption that $W[r]\neq \fpt$. 


An \emph{\fpt approximation scheme} (\fptas)\footnote{Not to be confused with the {\it fully polynomial-time approximation scheme} (\textup{FPTAS}).} is an approximation algorithm that, given an instance $I$ of a maximization problem $\Pi$, an integer $k$, and any $\epsilon > 0$, returns a $\frac{1}{(1+\epsilon)}$-approximate solution in $f(k, \epsilon)\cdot |I|^{\Oh(1)}$ time, where $f$ is an arbitrary computable function. It can be viewed as an \fpt algorithm with respect to parameters $\epsilon$ and $k$. For more details, we refer the reader to the texts by~\cite{ParamAlgorithms15b,FlumG06,downey2013fundamentals,DBLP:books/ox/Niedermeier06}.

\section{Obnoxious Egalitarian Chamberlin-Courant (CC)}\label{sec:egal-cc}

We begin our study with \OurCC. Recall that \problemshort with $\lambda=1$ is \OurCC. 
We begin with the Euclidean space, followed by the general metric space. 


\subsection{\texorpdfstring{Polynomial Time algorithm in $\B{R}^{2}$}{Polynomial Time algorithm in Rd for lambda = 1}}
%
%
In this section, we design a polynomial time algorithm when the voters and candidates are embedded in $\mathbb{R}^2$. In particular, we prove  the following result. 

\begin{theorem} \label{thm:polytime-plane}
There exists a polynomial-time algorithm to solve an instance of \OurCC when $\voters \cup \candidates \subset \Real^2$, and the distances are given by Euclidean distances.
\end{theorem}

\paragraph{Overview.} Before delving into a formal description of the polynomial-time algorithm, we start with a high-level overview of the result. For simplicity of the exposition, we assume that $t = 1$ (this can be easily achieved by scaling $\mathbb{R}^2$, and thus all points in the input by a factor of $t$). For each $c \in \candidates$, let $D(c)$ denote a \emph{unit disk} (i.e., an open disk of \emph{diameter} 1) with $c$ as its center. In the new formulation, we want to find a subset $S \subseteq \candidates$ of size $k$, such that for each $v \in \voters$, the solution $S$ contains at least one candidate $c$, such that $v$ is outside $D(c)$ (which is equivalent to saying that the euclidean distance between $v$ and $c$ is larger than $1$, which was exactly the original goal). This is an equivalent reformulation with a more geometric flavor, thus enabling us to use techniques from computational geometry. 

First, we perform some basic preprocessing steps, that will be help us in the main algorithm. First, if there is a disk $D(c)$ that does not contain a voter, then any set containing $c$ is a solution. Similarly, if we have two disjoint unit disks $D(c)$ and $D(c')$ centered at distinct $c, c' \in \candidates$, then any superset of $\left\{c, c'\right\}$ of size $k$ is a valid solution, which can be found and returned easily. We check this condition for subsets of size $2$, and also extend this check to subsets of size $3$. Now, assuming that the preprocessing step does not already give the solution, we know that each subset of unit disks of size at most $3$ have a common intersection. By a classical result in discrete geometry called Helly's theorem \cite{MatousekDG2002}, this also implies that each non-empty subset has a common intersection. Our goal is to find a smallest such subset $S$, for which, the common intersection region is devoid of all voters $v \in \voters$. We design a dynamic programming algorithm to find such a subset. Note that each subset is in one-to-one correspondence with a convex region defining the boundary of the common intersection, and the boundary of the common intersection consists of portions of boundaries of the corresponding unit disks (also known as ``arcs''). The dynamic programming algorithm considers partial solutions defined by a consecutive sequence of arcs that can be attached end-to-end, while at the same time, ensuring that the common intersection does not contain any voter $v \in \voters$. When we are trying to add another arc to the boundary, we have to make sure that (i) one of the endpoints of the arc is the same as one of the endpoints of the last arc defining the partial boundary, and (ii) the new area added to the ``partial common intersection'' does not contain a voter. We need to introduce several defintions and handle several special cases in order to formally prove the correctness of this strategy, which we do next. 

\paragraph{Formal description.} We work with the rescaled and reformulated version of the problem, as described above. Further, we assume, by infinitesimally perturbing the points if required (see, e.g., \cite{EdelsbrunnerM90}), that the points $\candidates \cup \voters$ satisfy the following general position assumption: no three unit disks centered at distinct candidates intersect at a common point.  Note that this assumption is only required in order to simplify the algorithmic description.

For each candidate $c \in \candidates$, let $D(c)$ denote the unit disk (i.e., an open disk of radius $1$) with $c$ as center. In the following, we will often omit the qualifier \emph{unit}, since all disks are assumed to be open unit disks unless explicitly mentioned otherwise. Note that our original problem is equivalent to determining whether there exists a subset $S \subseteq \candidates$ of size $k$ such that for every $v \in \voters$, there exists a candidate $c \in S$ such that $v \not\in D(c)$. Equivalently, we want to find a set $S \subseteq \candidates$ such that $\left(\bigcap_{c \in S} D(c)\right) \cap \voters = \emptyset$. For a subset $S' \subseteq\candidates$, we let $I(S') \coloneqq \bigcap_{c \in S'} D(c)$, and let $D(S') = \{D(c) : c \in S'\}$. We design a polynomial-time algorithm to find a smallest-sized subset $S' \subseteq \candidates$ such that $I(S') \cap \voters = \emptyset$. For any two points $x, y \in \Real^2$, let $\seg{xy}$ be the straight-line segment joining $x$ and $y$.

We first perform the following preprocessing steps to handle easy cases. For $k = 1$, we try each $c \in \candidates$ and check whether $d(v, c) \ge 1$ for all voters $v \in \voters$. Now suppose $k \ge 2$. First, we check whether there exists a pair of disks centered at distinct $c_1, c_2 \in \candidates$ such that the distance between $c_1, c_2$ is at least $2$. Then, for any voter $v \in \voters$, if $d(v, c_1) < 1$, then $d(v, c_2) > 1$ by triangle inequality. Therefore, $\left\{c_1, c_2\right\}$ can be augmented by adding arbitrary set of $k - 2$ candidates in $\candidates \setminus \left\{c_1, c_2\right\}$ to obtain a solution. Now suppose that neither of the previous two steps succeeds. Then, we try all possible subsets $S' \subseteq \candidates$ of size at most $3$, and check whether $I(S') = \emptyset$, that is, no point in $\Real^2$ belongs to $I(S')$ (note that this specifically implies that $I(S') \cap \voters = \emptyset$). If we find such a set $S'$, then we can add an arbitrary subset of $\candidates \setminus S'$ of size $k - |S'|$ to obtain a set $S$ of size $k$. 
Thus, we can make the following assumptions, given that the preprocessing step does not solve the problem.

\begin{enumerate}
	\item $k \geq 4$,
	\item For every $c, c' \in \candidates$, $D(c) \cap D(c') \neq \emptyset$, and the two disks intersect at two distinct points (this is handled in the second step of preprocessing), and
	\item For any subset $\emptyset \neq S \subseteq \candidates$, $I(S) \neq \emptyset$. In particular, this also holds for sets $S$ with $|S| > 3$ -- otherwise by Helly's theorem \cite{MatousekDG2002}, 
 there would exist a subset $S' \subseteq S$ of size $3$ such that $I(S') = \emptyset$, a case handled in the preprocessing step.
\end{enumerate}

Let $\cP$ be a set of intersection points of the boundaries of the disks $\{D(c): c \in \candidates\}$. Note that since the boundaries of every pair of disks intersect exactly twice (this follows from the item (2) above), $|\cP| = 2 \binom{|\candidates|}{2}$ . Furthermore, for $c \in \candidates$, let $\cP(c) \subset \cP$ be the set of intersection points that lie on the boundary of $D(c)$. For $c \in \candidates$ and distinct $p, q \in \cP(c)$, we define $\arc(p, q, c)$ as the \emph{minor arc} (i.e., the portion of the boundary of $D(c)$ that is smaller than a semicircle) of disk $D(c)$ with $p$ and $q$ as its endpoints. Note that $p$ and $q$ are interchangeable in the definition, and $\arc(p, q, c) = \arc(q, p, c)$. For a subset $S' \subseteq \candidates$, let $\cA(S')$ be the set of arcs defining the boundary of the region $I(S')$ -- note that since $I(S') \neq \emptyset$ for any $S' \neq \emptyset$, $\cA(S')$ is well-defined and is a non-empty set of arcs. We first have the following proposition, the proof of which follows from arguments in planar geometry.


\begin{proposition} \label{prop:geom}
	Fix a set $S \subseteq \candidates$ with $|S| \ge 2$. Furthermore, assume that $S$ is a \emph{minimal} set with intersection equal to $I(S)$, i.e., there exists no subset $S' \subset S$ such that $I(S') = I(S)$.	Then, for every $c \in S$, $\cA(S)$ contains exactly one arc of the form $\arc(p, q, c)$ for some $p, q \in \cP(c)$. 
\end{proposition}
\begin{proof}
	First we prove that every arc in $\cA(S)$ is a minor arc. Suppose for contradiction that $\cA(S)$ contains a non-minor arc $A$ on the boundary of some $D(c)$, $c \in S$. Consider any $c' \in S$ with $c' \neq c$, and let $S' = \{c, c'\}$. Note that $I(S) \subseteq I(S')$ as $S'\subseteq S$ and intersection of disk can only decrease by adding more points to the set. 
 Thus, $\cA(S')$ contains an arc $A'$ that is a superset of $A$. Let $p$ and $q$ denote the endpoints of $A'$, and note that $A'$ is also a non-minor arc. Note that $p \in I(S') = D(c) \cap D(c')$. Let $p'$ denote the point on $D(c)$ that is diametrically opposite to $p$, and since $A'$ is a major arc, it follows that $p' \in A' \subseteq I(S') = D(c) \cap D(c')$. To summarize, both $p$ and $p'$ belong to both $D(c)$ and $D(c')$. However, since both $D(c)$ and $D(c')$ are unit disks, $\seg{pp'}$ 
 is a common diameter of $D(c)$ and $D(c')$, which contradicts that $c$ and $c'$ are distinct. 
	
	Now we prove the second part of the claim, that is, for each $c \in S$, $\cA(S)$ contains exactly one minor arc of the form $\arc(\cdot, \cdot, c)$. Suppose there exists some $c$ such that there exist two arcs $A_1 = \arc(p_1, q_1, c)$ and $A_2 = \arc(p_2, q_2, c)$ in $\cA(S)$. Note that $A_1$ and $A_2$ must be disjoint, otherwise we can concatenate them to obtain a single arc. Suppose, without loss of generality, traversing clockwise along the boundary of $D(c)$, the ordering of the points is $p_1, q_1, q_2, p_2$. Let $c_1 \in S$ (resp.\ $c_2 \in S$) be the candidate such that $q_1$ (resp.\ $q_2$) belongs on the boundaries of $D(c)$ and $D(c_1)$ (resp.\ $D(c)$ and $D(c_2)$). It is clear that $c \neq c_1$ and $c \neq c_2$. Now, we additionally claim that $c_1 \neq c_2$ -- suppose this is not the case. Then, $q_1$ and $q_2$ belong to the boundaries of $D(c)$ and $D(c_1)$. In this case, $p_1$ (or $p_2$) cannot belong to $D(c) \cap D(c_1) \subseteq I(S)$, 
 which contradicts the assumption that $p_1$ (or $p_2$) lie on the boundary of $I(S)$. Thus, we have that $c, c_1, c_2$ are all distinct. However, again we reach a contradiction since $p_1$ is outside $D(c) \cap D(c_2) \subseteq I(S)$. Thus, it follows that each arc appears at most once in $\cA(S)$. 
	
	Finally, we consider the case when there exists some $c \in S$ such that no arc of the form $\arc(\cdot, \cdot, c)$ belongs to $\cA(S)$. In this case, the region bounded by $\cA(S)$, i.e., $I(S)$, is completely contained inside $D(c)$. However, this implies that $I(S) =\! I(S \setminus\! \{c\})$, which contradicts the minimality of $S$.  
\qed \end{proof}

\begin{figure*}
	\centering
	\includegraphics[scale=0.5]{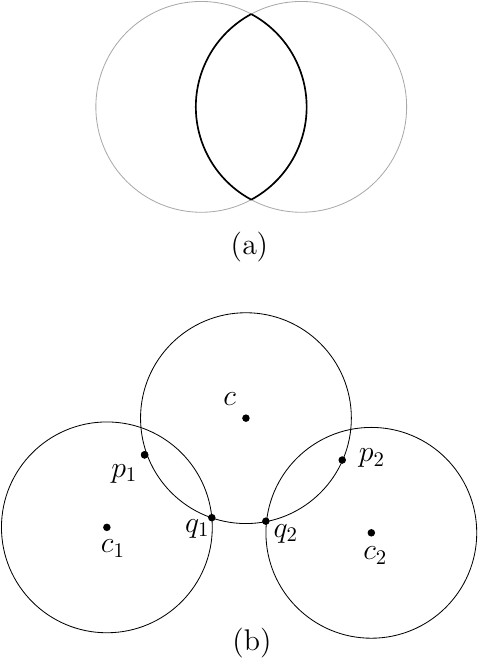}\hspace{1cm}
	\includegraphics[scale=1]{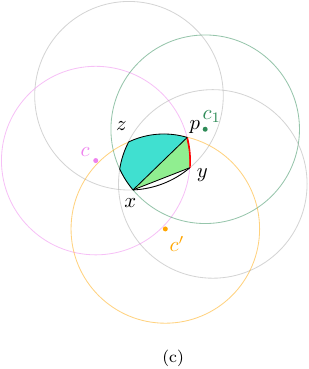}
	\caption{Illustration for the proof of Proposition \ref{prop:geom} and the algorithm. Fig (a): intersection of boundaries of two unit disks is defined by two minor arcs. Fig (b): Two disjoint arcs $\arc(p_1, q_1, c)$ and $\arc(p_2, q_2, c)$ cannot appear on the boundary of a common intersection, since they correspond to disjoint regions. Fig (c): Illustration for the dynamic program. A region formed by intersection of $5$ disks is shown. $\arc(p, y, c)$ is shown in red. Blue region corresponds to the entry $A[x, p, z, c_1, c', 3]$, and green region corresponds to the newly added region to the blue region, corresponding to the entry $A[x, y, p, c_1, c, 4]$.}\label{figure1}
\end{figure*}

Next, we proceed towards designing our algorithm. 
\paragraph{\bf Algorithm.}\label{para:algorithm}
 Now, we proceed to the dynamic programming algorithm. 
For any $x, y, p \in \cP$, $c_1, c \in \candidates$, and an integer $i \ge 2$, we define a table entry $A[x, y, p, c_1, c, i]$ that denotes whether there exists a region $R(x,y,p,c_1, c) \subset \Real^2$ with the following properties:
\begin{itemize}[wide=0pt]
	\item $R = R(x,y,p,c_1,c)$ is a convex region bounded by a set $\cA(R)$ of $i-1$ circular arcs, and straight-line segment $\seg{xy}$, such that $\cA(R)$ contains: 
	\begin{itemize}
		\item At most one arc of the form $\arc(\cdot, \cdot, c')$ for every $c' \in \candidates$.
		\item Exactly one arc of the form $\arc(x, \cdot, c_1)$, which is the \emph{first} arc traversed along the boundary of $R$ in clockwise direction, starting from $x$. Note that $c_1$ is the center of this arc.
		\item $\arc(y, p, c)$ 
	\end{itemize}
	\item $R \cap \voters = \emptyset$.
\end{itemize}
Note that if $\arc(y, p, c)$ is not defined, or any of the other conditions do not hold, then a region with the required properties does not exist. 

First, we compute all entries $A[x, y, p, c, c_1, i]$ with $i \le 3$. Note that the number of arcs of the form $\arc(\cdot, \cdot, \cdot)$ is bounded by $O(|\cP|^2 \cdot |\candidates|) = O(|\candidates|^3)$, and since $i \le 3$, we can explicitly construct all such candidate regions in polynomial time. Thus, we can correctly populate all such table entries with \btrue or \bfalse.

Now, we discuss how to fill a table entry $A[x, y, p, c_1, c, i]$ with $i \ge 4$. We fix one such entry and its arguments. If the region $R(x, p, y, c_1, c)$ 
bounded by $\seg{xp}, \seg{xy}$ and $\arc(y, p, c)$ contains a point from $\voters$, then the entry $A[x,p, y,c_1, c, i]$ is defined to be \texttt{false}. Note that this can easily be checked in polynomial time. Otherwise, suppose that $R(x, p, y, c_1, c) \cap \voters = \emptyset$. In this case, let $\mathcal{T}$ be a set of tuples of the form $(x, p, z, c_1, c', i-1)$, where $z \in \cP$, and $c' \in \candidates$ such that the following conditions are satisfied. (1) $c' \not\in \{c, c_1\}$, (2) The minor arc $\arc(p, z, c')$ exists, and (3) When traversing along this arc from $z$ to $p$, the arc $\arc(p, y, c)$ is a ``right turn''. Formally, consider the tangents $\ell_{c}, \ell_{c'}$ to the disks $D(c)$ and $D(c')$ at point $p$ respectively. Let $H_c$ (resp. $H_{c'}$) be the closed halfplane defined by the line $\ell_c$ ($\ell_{c'}$) that contains $D(c)$ ($D(c')$). Then, arcs $\arc(p, z, c')$ and $\arc(y, p, c)$ must belong to $H_{c} \cap H_{c'}$. 
See Figure~\ref{figure1}(c). Then,
\begin{align*}
	A[x, y, p, c_1, c, i] &= \bigvee_{(x, p, z, c_1, c', i-1) \in \mathcal{T}} A[x, p, z, c_1, c', i-1].
\end{align*}
Since we take an or over at most $|\candidates| \times |\voters|$ many entries, each such entry can be computed in polynomial time. Furthermore, since the number of entries is polynomial in $|\candidates|$ and $|\voters|$, the entire table can be populated in polynomial time.

Now, we iterate over all entries $A[x, y, p, c_1, c, i]$ such that the following conditions hold.
\begin{itemize}[wide=0pt]
	\item $A[x, y, p, c_1, c, i] = \btrue$, 
	\item There exists some $c'' \in \candidates \setminus \left\{c, c_1\right\}$ such that $\arc(x, y, c'')$ exists, and
	\item The region bounded by $\arc(x, y, c'')$ and segment $\seg{xy}$ does not contain any point from $\candidates$. The meaning here is that $\arc(x, y, c'')$ is the last arc bounding the required region. 
\end{itemize}
If such an entry exists with $i \le k-1$, then we conclude that the given instance of \problemshort is a yes-instance. Otherwise, it is a no-instance. Finally, using standard backtracking strategy in dynamic programming, 
it actually computes a set $S \subseteq \candidates$ such that $I(S) \cap \voters = \emptyset$. 

Next, we establish the proof of correctness of this dynamic program.

\paragraph{\bf A Proof of Correctness.}

\begin{lemma}($\clubsuit$\footnote{The proofs marked with $\clubsuit$ can be found in Appendix.}) \label{lemma:dp-correctness}
	Consider an entry $A[x, y, p, c_1, c, i]$, corresponding to some $x, y, p \in \cP, c_1, c \in \candidates$. Then, $A[x, y, p, c_1, c, i] = \btrue$ if and only if the corresponding region $R$, as in the definition of the table entry, contains no point of $\voters$.
\end{lemma}

\begin{lemma}
	This algorithm correctly decides whether the given instance of \problemshort is a \yes-instance. 
\end{lemma}
\begin{proof}
	First, it is easy to see that the preprocessing step correctly finds a minimum-size subset of at most $3$ whose intersection contains no point of $\voters$, if such a subset exists. Thus, we now assume that the preprocessing step does not find a solution, and the algorithm proceeds to the dynamic programming part.
	
	Recall that due to Proposition 1, if $S$ is a minimal subset of centers, such that $I(S) \cap \voters = \emptyset$ (if any), then $\cA(S)$ contains exactly one minor arc that is part of the circle centered at each $c \in S$. In particular, this holds for the optimal set $S^*$ of centers (if any), and let $i = |S^*|$. Pick an arbitrary arc in $\cA(S^*)$, and let $c_1$ be the center of this arc, and $x$ be one of the endpoints of this arc. By traversing the arcs in $\cA(S^*)$ in clockwise manner, let the last two arcs be $\arc(y, p, c)$, and $\arc(y, x, c')$. Then, by \Cref{lemma:dp-correctness}, it follows that $A[x, y, p, c_1, c', i-1] = \btrue$, and the region bounded by $\seg{xy}$ and $\arc(y, x, c')$ does not contain a point from $\candidates$. Thus, the algorithm outputs the correct solution corresponding to the entry $A[x, y, p, c_1, c', i-1]$. 
	
	In the other direction, if the algorithm finds an entry $A[x, y, p, c_1, c', i-1] = \btrue$, such that (1) $\arc(x, y, c)$ exists, (2) $c \not\in \{c', c_1\}$, and (3) the region bounded by $\arc(x, y, c)$ and $\seg{xy}$ does not contain a point from $\candidates$, then using \Cref{lemma:dp-correctness}, we can find a set of $i$ disks whose intersection does not contain a point from $\candidates$. Therefore, if for all entries it holds that at least one of the conditions does not hold, then the algorithm correctly concludes that the given instance is a no-instance.
\qed \end{proof}

The algorithm as it is does not work for $\lambda>1$.
We do not know whether the problem is polynomial time solvable for $\lambda>1$. 



\subsection{Hardness in Graph Metric}\label{sec:nph-metric-lambda1}

In this section, we show the intractability of the problem when the voters and candidates are embedded in the graph metric space, which implies the intractability in the general metric space. The metric space defined by the vertex set of a graph as points and distance between two points as the shortest distance between the corresponding vertices in the graph is called the {\it graph metric space}. 

We present a reduction from the {\sc Hitting Set} problem, defined below, which is known to be \nph \cite{Karp72} and W$[2]$-parameterized by $k$ \cite{ParamAlgorithms15b}.

\defproblem{{\sc Hitting Set}}{Set system $(\cU, \cF)$, where $\cU$ is the ground set of $n$ elements, $\cF$ is a family of subsets of $\cU$, and a positive integer $k$}
{Does there exist $H \subseteq \cU$ of size $k$ such that for any $S \in \cF$, $H \cap S \neq \emptyset$?}

\noindent{\bf Reduction:} Define a graph $G$ with vertex set $\voters \cup \candidates$ as follows. For every element $e \in \cU$, we add a candidate $c_e$ to $\candidates$, and for every set $S \in \cF$, we add a voter $v_S$ to $\voters$. We add an edge $(c_e, v_S)$  in $G$ if and only if $e \not\in S$. The weight of all such edges is equal to $1$. Also, for any $c_e, c_e' \in \candidates$, we add an edge of weight $2$. The distance function $d: (\voters \cup \candidates) \to \mathbb{R}^+$ is given by the shortest path distances in $G$. 


\begin{observation} \label{obs:hittingset-equiv}
	For any $e \in \cU$, and $S \in \cF$, $d(c_e, v_S) = 1$ if and only if $e \not\in S$. Otherwise $d(c_e, v_S) = 3$ if and only if $e \in S$. 
\end{observation}

\begin{lemma}($\clubsuit$)\label{lem:hittingset-equiv}
	$(\cU, \cF)$ admits a hitting set of size $k$ if and only if there exists a set $H \subseteq \candidates$ of size $k$ such that for any $v_S \in \voters$, $\max_{c_e \in H} d(c_e, v_S) = 3$. 
\end{lemma}

\hide{
\begin{proof}[Proof of Lemma~\ref{lem:hittingset-equiv}]
	In the forward direction, let $H\subseteq \cU$ be a hitting set of size at most $k$. By adding arbitrary elements, we obtain $H' \subseteq \cU$ of size exactly $k$, which remains a hitting set. Let $H''$ be the subset of $\candidates$ corresponding to $H'$. Now, for any $S \in \cF$, $H$ contains an element $e$ such that $e \in S$. Thus, for every $v_S \in \voters$, there exists a $c_e \in H''$, such that $d(c_e, v_S) = 3$ by Observation \ref{obs:hittingset-equiv}. 
	
	In the reverse direction, let $H \subseteq \candidates$ be a set of size $k$ such that for any $v_S \in \voters$,  $\max_{c_e \in H} d(c_e, v_S) = 3$. Let $H'$ be the corresponding subset of elements. 
	For every $v_S \in \cF$, $H$ contains a candidate $c_e$ such that $d(c_e, v_S) = 3$. By Observation \ref{obs:hittingset-equiv}, $e \in S$. This implies that $H'$ is a hitting set.
\qed \end{proof}}
In fact, this construction shows that it is NP-hard to approximate the problem within a factor of $1/3 + \epsilon$ for any $\epsilon > 0$. Indeed, suppose there existed such a $\beta =(1/3 + \epsilon)$-approximation for some $\epsilon > 0$. Then, if $(\cU, \cF)$ is a yes-instance of \textsc{Hitting Set}, then Lemma \ref{lem:hittingset-equiv} implies that $\textsf{OPT} = 3$ -- here \textsf{OPT} denotes the largest value of $t$ for which we have a yes-instance for the decision version. In this case, the $\beta$-approximation returns a solution $S$ of size $k$ and of cost at least $\beta \cdot 3 = 1 + 3 \epsilon > 1$. This implies that for each $v_S \in \voters$, there exists some $c_e \in S$ with $d(v_S, c_e) > 1$. However, such a $c_e$ must correspond to an element $e \in S$ -- otherwise $d(v_S, c_e) = 1$ by construction. Therefore, the solution $S$ corresponds to a hitting set of size $k$. Alternatively, if $(\cU, \cF)$ is a no-instance of \textsc{Hitting Set}, then Lemma \ref{lem:hittingset-equiv} implies that the there is no solution of size $k$ with cost $3$. Thus, a $\beta$-approximation can be used to distinguish between yes- and no-instances of \textsc{Hitting Set}. 
Hence, we have the following result. 

\begin{theorem} \label{thm:apx-hardness-cc}
	 \OurCC is \nph. Furthermore, for any $\alpha > 1/3$, \OurCC does not admit a polynomial time $\alpha$-approximation algorithm, unless $\mathsf{P = NP}$. Furthermore, \OurCC does not admit an \fpt-approximation algorithm parameterized by $k$ with an approximation guarantee of $\alpha \ge 1/3$, unless $\fpt = W[2]$.
\end{theorem}






\subsection{Approximation Algorithm in General Metric Space}

In this section, we design a polynomial time  $1/4$-approximation algorithm when voters and candidates are embedded in a general metric space. 

We first guess a voter $p' \in \voters$ and a candidate $c' \in \candidates$ such that $c'$ is the farthest candidate from $p'$ in an optimal solution $S^*$. Let $t = d(p', c')$. We know that all candidates in $S^*$ are within a distance $t$ from $p'$, i.e., $S^* \subseteq B$, where $B = B(p', t) \cap \candidates$. 
Let $M$ be a $(t/2)$-net of $B$, 
i.e., $M$ is a set of candidates with following properties: (1) $d(c_i, c_j) > t/2$ for any distinct $c_i, c_j \in M$, and (2) for any $c \not\in M$, there exists some $c' \in M$ such that $d(c, c') \le t/2$. 

Now there are two cases. 
(1) If $|M| \ge k$, let $M'$ be an arbitrary subset of $M$ of size exactly $k$. 
(2) If $|M| < k$, then let $M' = M \cup Q$ where $Q$ is an arbitrary subset of candidates from $B \setminus M$ such that $|M'| = k$.

\begin{lemma}
	Fix some $v \in \voters$, and let $c \in \candidates$ 
 be the farthest center from $v$ in $M'$. Then, $d(v, c) \ge t/4$.
\end{lemma}
\begin{proof}
	We consider two cases based on the size of $M$, and the way we obtain $M'$ from $M$.
	
	Case 1: $|M| \ge k$, and $M'$ is an arbitrary subset of $M$.
	\\Suppose for contradiction $d(v, c) < t/4$. Since $c$ is the farthest candidate from $v$, 
 the same is true for any candidate $c' \in M$. Then, $d(p, c) < t/4$ and $d(v, c') < t/4$, which implies that $d(c, c') \le d(v, c) + d(v, c') < t/4 + t/4 = t/2$, which contradicts property 1 of $M$. 
	
	Case 2. $|M| < k$ and $M'$ is obtained by adding arbitrary candidates to $M$.
	\\If $d(v, c) \ge t/2 \ge t/4$, we are done. So assume that $d(p, c) < t/2$.
	\\Let $c^*$ be the farthest center from $v$ in an optimal solution. Then, $d(v, c^*) \ge t$. Also, $c^* \not\in M \subseteq M'$, otherwise $d(v, c) \ge d(v, c^*) \ge t$, since $c$ is the farthest candidate from $v$. Therefore, by property 2, there exists some $c' \in M$ such that $d(c', c^*) \le t/2$. Again, $d(v, c') \le d(v, c) < t/2$. Then, $d(v, c^*) \le d(v, c') + d(c', c^*) < t/2 + t/2 = t$. 
 This contradicts $d(v, c^*) \ge t$.
\qed \end{proof}
Thus, we conclude with the following theorem.
\begin{theorem}\label{thm:polytime-approx}
	There exists a polynomial-time $1/4$-approximation algorithm for the optimization variant of \OurCC when the voters and candidates belong to an arbitrary metric space $\X{M}$.
\end{theorem}


\section{\problem for $\lambda>1$}\label{sec:large value}

In this section, we move our study to the case when $\lambda>1$. We first show that $\lambda=k-1$ is \nph in Euclidean space. But the extreme cases of $\lambda=1$ or $\lambda=k$ are tractable: Infact, the $\lambda=1$ case is polynomial-time solvable for Euclidean space but the $\lambda=k$ is polynomial-time solvable even for a general metric space. Furthermore, we show that similar to $\lambda=1$, the problem is hard to approximate for any value of $\lambda$ in graph metric.  Finally, contrary to Theorem~\ref{thm:apx-hardness} we give an {\sf FPT}-approximation scheme for arbitrary value of $\lambda$ in Euclidean space. 

\subsection{Hardness}
In this section, we present results pertaining to {\sf NP}-hardness and approximation hardness.
\paragraph{\bf NP-hardness for $\lambda=k-1$ in $\mathbb{R}^2$.}


To exhibit this we give a reduction from the \textsc{2-Independent Set} problem in unit disk graphs (UDGs). We give a formal definition of UDGs below, followed by the definition of the aforementioned problem.

\begin{definition}Given a set $\X{P} = \{p_{1}, p_{2}, \ldots,p_{n}\}$ of points in the plane, a unit disk graph (UDG, in short) corresponding to the point set \X{P} is a simple graph $G = (\X{P},E)$ satisfying $E =\{(p_{i}, p_{j} ) | d(p_{i}, p_{j} ) \leq 1\}$, where $d(p_{i}, p_{j})$ denotes the Euclidean distance between $p_{i}$ and $p_{j}$ .
\end{definition}


\defproblem{2-{\sc Independent Set In Unit Disk Graph}}{Given a set $V \subset \Real^2$ of $n$ points, and a positive integer $k$.}{Let $G = (V, E)$ be a unit disk graph 
defined on $V$. Does there exist a subset $S\sse V$ such that $|S| = k$, and for any distinct $u, w \in S$, $d_G(u, w) > 2$?}

This problem is shown to be \nph in \cite{jena2018maximum}.

\begin{theorem} \label{thm:r2-hardness}
	\problemshort is \nph when $\lambda= k-1$, even in the special case where $\voters \cup \candidates \subset \Real^2$ and the distances are given by standard Euclidean distances.
\end{theorem}
\begin{proof}
	Let $(V, k)$ be the given instance of 2-{\sc Independent Set In Unit Disk Graph}, where $V \subset \Real^2$. We create an instance of \problemshort as follows. For every point $p \in V$, we add a voter and a candidate co-located at the point in $\Real^2$ at the point $p$. Let $\voters$ and $\candidates$ be the resulting sets of $n$ voters and $n$ candidates, and the value of $k$ remains unchanged. Without loss of generality, we assume that $k \ge 2$. We set $\lambda = k-1$. We prove the following lemma. Note that due to the strict inequality, this does not quite fit the definition of \problemshort. Subsequently, we discuss how to modify the construction so that this issue is alleviated. In the following proof, we use $d_e(\cdot, \cdot)$ to denote the Euclidean distance and $d_G(\cdot, \cdot)$ to denote the shortest-path distance in the unit disk graph $G$.
	
	\begin{lemma} \label{lem:equiv-r2-hardness}
		$S$ is a $2$-independent set of size $k$ in $G$ if and only if for the corresponding set $S' \subseteq \candidates$, it holds that, for every voter $v \in \voters$, $d_e^{\lambda}(v, S') > 2$.
	\end{lemma}
	\begin{proof}
		In the forward direction, let $S$ be a $2$-independent set of size $k$ in $G$, and let $S'$ be as defined above. Suppose for the contradiction that there exists a voter $v$ for which $d_e^{\lambda}(v, S') \le 2$. That is, there exists two distinct candidates $c_1, c_2 \in S'$ such that $d_e(v, c_1) \le 2$, and $d_e(v, c_2) \le 2$. We consider two cases, depending on whether $v$ is co-located with either of $c_1$ or $c_2$, or not. Suppose $v$ is co-located with $c_1$ (w.l.o.g., the $c_2$ case is symmetric). Then, let $p_1$ and $p_2$ be the points in $S \subseteq P$ corresponding to $v$ and $c_2$ respectively. However, since $d_e(p_1, p_2) \le 2$, $(p_1, p_2)$ is an edge in $G$, which contradicts the $2$-independence of $S$. In the second case, $v$ is not co-located with $c_1$ as well as $c_2$. Even in this case, let $q, p_1, p_2$ be the points in $P$ corresponding to $v$, $c_1$, and $c_2$ respectively. Note that $q, p_1, p_2$ are distinct, and $p_1, p_2 \in S$. However, since $d_e(q, p_1) \le 2, d_e(q, p_2) \le 2$, $(q, p_1)$ and $(q, p_2)$ are edges in $G$, which again contradicts the $2$-independence of $S$, as $d_G(p_1,p_2)=2$.
		
		In the reverse direction, let $S' \subseteq \candidates$ be a subset of candidates such that for each voter $v \in \voters$, $d_e^{\lambda}(v, S') > 2$. Let $S \subseteq P$ be the corresponding points of $S'$, and suppose for contradiction that $S$ is not a $2$-independent set in $G$, which implies that there exist two distinct $p_1, p_2 \in S$ such that $d_G(p_1, p_2) \le 2$. Let $c_1, c_2$ be the candidates in $S'$ corresponding to $p_1$ and $p_2$ respectively. Again, we consider two cases. First, suppose that $d_G(p_1, p_2) = 1$, i.e., $(p_1, p_2)$ is an edge in $G$. Then, let $v_1$ be the voter co-located at $p_1$. Then, for $v_1$, $d_e(v_1, c_1) = 0$, and $d_e(v_1, c_2) \le 2$, since $(p_1, p_2)$ is an edge. This contradicts that $d_e^{\lambda}(v_1, S') > 2$. In the second case, suppose $d_G(p_1, p_2) = 2$, then let $q \in P$ be a common neighbor of $p_1$ and $p_2$ in $G$, and let $v_q \in \voters$ be the voter co-located to $q$. Again, note that $d_e(v_q, c_1) \le 2$ and $d_e(v_q, c_2) \le 2$, which contradicts that $d_e^{\lambda}(v_q, S') > 2$. 
	\qed \end{proof}
 
	Let $\displaystyle t \coloneqq \min_{p, q \in P: d_e(p, q) > 2} d_e(p, q)$. That is, $t$ is the smallest Euclidean distance between non-neighbors in $G$. By definition, for any $p', q' \in P$ such that $d_e(p', q') > 2$, it holds that $d_e(p', q') \ge t$. Now, we observe that the proof of Lemma \ref{lem:equiv-r2-hardness} also works after changing the condition $d_e^{\lambda}(v, S') > 2$ to $d_e^{\lambda}(v, S') \ge t$. Note that there exists points $p,q$ such that $d_G(p,q)>2$, and hence $d_e(p, q) > 2$,   otherwise, it is a trivial no-instance of 2-{\sc Independent Set In Unit Disk Graph}.
\qed \end{proof}

\if and only ifalse
\defproblem{2-{\sc Independence In UDG}}{Given a UDG $G=(V,E)$ and an integer $k$}{Does there exist a subset $S\sse V$ such that $|S| \leq k$ and no pair of vertices in $S$ induces a $P_{2}$ ?}

\begin{theorem}
	\problemshort is \nph when $\lambda= k-1$.
\end{theorem}

\begin{proof}We give a reduction from 2-{\sc Independence In UDG} as follows. 
	\Ma{The main idea behind this ....}
	
	Let $\Co{I}=(G, k)$ is an instance of 2-{\sc Independence In UDG}. We will create an instance of \problemshort, $\Co{J}= (\X{M}, d, t, \lambda)$, where \X{M} is a metric space containing points from the set \voters and \candidates and $d(\cdot, \cdot)$ is the distance metric; while $t=2$ and $\lambda =k-1$.
	
	
	\noindent{\bf Reduction:}
	For every vertex $v\in V(G)$, we have two points, $v_{\vp} \in \voters$ and $v_{\cp} \in \candidates$. For every pair $(a_{\vp}, b_{\vp}) \in \voters \times \candidates$, we define the distance as follows, where $0<\epsilon < 1/2$   
	\[ d(a_{\vp}, b_{\cp}) = \begin{cases} t- \epsilon & \textrm{ if } (a,b) \in E(G);\\
		t+\epsilon & \textrm{ if } (a,b) \notin E(G);\\
	\end{cases}
	\]
	
	
	Additionally, for all other pair of points the distance is $t+2\epsilon$. 
	
	
	\begin{clm}
		The triangle inequality holds in \X{M}. 
	\end{clm}
	
	\begin{proof}Consider any three arbitrary points $p_{1}, p_{2}, p_{3}\in \X{M}$. We observe that $d(p_{1}, p_{2}) + d(p_{2}, p_{3}) \geq 2(t-\epsilon)$ and $d(p_{1}, p_{3}) \leq t + 2\epsilon$. Since $0 <\epsilon <1/2$, we have  $d(p_{1}, p_{2}) + d(p_{2}, p_{3}) > d(p_{1}, p_{3})$. 
		
		\begin{suppress}
			To explain more details, we note that the analysis can be broken into following cases. 
			
			Let $p_{1} \in \candidates$. Then, we can only the following two possibilities
			\begin{itemize}
				\item Suppose both $p_{2}, p_{3}\in \voters$. \hide{Then, $d(p_{1}, p_{2}) \geq t-\epsilon$, $d(p_{2}, p_{3}) = t+2\epsilon$ and $d(p_{1}, p_{3}) \leq t + \epsilon$.} It follows that $d(p_{1}, p_{2})+ d(p_{2}, p_{3}) \geq 2t +\epsilon >t + \epsilon \geq  d(p_{1}, p_{3})$. 
				
				\item Suppose both $p_{2}, p_{3} \in \candidates$. Since, $d(p_{i}, p_{j}) =t + 2\epsilon$
				for all $\{i, j\} \sse \{1,2,3\}$, we have $ d(p_{1}, p_{2})+ d(p_{2}, p_{3}) > d(p_{1}, p_{3})$. 
				
				\item Suppose $p_{2} \in \voters$ and $p_{3}\in \candidates$. \hide{Since $d(p_{1}, p_{3}) = t+2\epsilon$, $d(p_{1}, p_{2}), d(p_{2}, p_{3})\geq t-\epsilon$.} Hence, $d(p_{1}, p_{2})+ d(p_{2}, p_{3}) \geq 2(t- \epsilon) > d(p_{1}, p_{3})= t + 2\epsilon$.

				\item Suppose that $p_{2}\in \candidates$ and $p_{3}\in \voters$. \hide{Then $d(p_{1}, p_{2}) = t+2\epsilon$, $d(p_{2}, p_{3}) \geq t-\eps$, and $d(p_{1}, p_{3}) \leq t+\epsilon$.} Hence, $d(p_{1}, p_{2}) + d(p_{2}, p_{3}) \geq 2t + \epsilon > t+ \epsilon \geq d(p_{1}, p_{3})$. 
			\end{itemize}
			We can similarly analyse the case when $p_{1} \in \voters$. 
		\end{suppress}

	\qed \end{proof}

	%
	%
	
	Let $S$ be a solution of \Co{I}. We define $\hat{S}_{\cp}=\{v_{\cp} \in \candidates: v\in S\}$. 

	\begin{clm}
		The subset $\hat{S}$ is a solution for \Co{J}. 
	\end{clm}
	
	\begin{proof}We need to show that for every point $p_{\vp} \in \voters$, there exist at least $k-1$ points in $\hat{S}$ that are ``far'' (i.e distance $t$ away) from $p$. We note that if there exists a point $a_{\cp}\in \hat{S}$ such that $d(a_{\cp}, p_{\vp}) < t$, then the corresponding vertices $\{p, a\} \sse V(G)$ must be adjacent. Thus, for any other points $b_{\cp} \in \hat{S} \sm \{a_{\cp}\}$, we know that the corresponding vertex $b \in S$ must not be adjacent to either $p$ or $a$. Thus, $d(b_{\cp}, a_{\vp}), d(b_{\cp}, p_{\vp})> t$. Hence, $\hat{S}$ is indeed a solution for \Co{J}. 
	\qed \end{proof}


	For the other direction, let $\hat{S}$ denote a solution for \problemshort. Clearly, $\hat{S} \sse \candidates$. Consider the set $S = \{x\in V(G): x_{\cp} \in \hat{S}\}$.
	
	\begin{clm}
		Subset $S$ is a solution for \Co{I}.
	\end{clm}
	
	\begin{proof} Consider any two vertices $x$ and $z$ in $S$. Clearly, the \candidates-points $x_{\cp}, z_{\cp} \in \hat{S}$. Suppose that $x$ and $z$ have a common neighbor $y$ in $G$. Then, $y_{\vp} \in \voters$ and $d(x_{\cp}, y_{\vp}) = d(y_{\vp} ,z_{\cp}) = t - \epsilon$. Thus, there are two points in $\hat{S}$ that are ``close'' to the \voters-point $y_{\vp}$, a contradiction.
	\qed \end{proof}
	
	This concludes the proof. 
\qed \end{proof}
\fi



\paragraph{\bf Approximation Hardness in Graph Metric}
The reduction is same as in Section~\ref{sec:nph-metric-lambda1}. Here, instead of {\sc Hitting Set}, we give a reduction from the {\sc Multi-Hitting Set} problem, where each set needs to be hit at least $\lambda \ge 1$ times for some constant $\lambda$. It can be easily seen that this is a generalization of {\sc Hitting Set} and is also NP-complete \cite{RajagopalanV98} (for an easy reduction, simply add $\lambda-1$ ``effectively dummy'' sets that contain all the original elements)
 Thus, we have the following result. 

\begin{theorem} \label{thm:apx-hardness}
	For any fixed $1 \le \lambda < k$, \problemshort is \nph. Furthermore, for any fixed $1 \le \lambda < k$, and for any $\alpha \ge 1/3$, \problemshort does not admit a polynomial time $\alpha$-approximation algorithm, unless $\mathsf{P = NP}$. Furthermore, \problemshort does not admit an FPT-approximation algorithm parameterized by $k$ with an approximation guarantee of $\alpha \ge 1/3$, unless $FPT = W[2]$.
\end{theorem}

\subsection{FPT-AS in Euclidean and Doubling Spaces}

In this section, we design an {\sf FPT} approximation scheme for the inputs in $\Real^d$, parameterized by $\lambda, d$, and $\epsilon$. In fact, the same arguments can be extended to metric spaces of doubling dimension $d$. However, we focus on $\Real^d$ for the ease of exposition, and discuss the case of doubling spaces at the end.

In the subsequent discussions, we say that $S \subseteq \candidates$ is a \emph{solution} if it satisfies the following two properties: (i) $|S| \ge \lambda$, and (ii) for each $v \in \voters$, $d^{\lambda}(v, c) \ge t$. For any given instance of \problemshort, we note the following simple observation. 

\begin{observation}
	If there exists $S\sse \candidates$ of size at least $\lambda$, such that each $c \in S$ is at distance at least $t$ from each $v\in \voters$, then $S$ is a solution. 
\end{observation}


Moreover, we can infer the following based on the pairwise distances between the points in \candidates.

\begin{observation}\label{obs1}
	A subset $S \sse \candidates$ of $\lambda+1$ points that are pairwise $2t$ distance away from each other is a solution. 
\end{observation}

\begin{proof}Let $S=\{c_{1}, \ldots, c_{\lambda+1}\}$ denote the aforementioned subset of points. 
	Consider any arbitrary point $v\in \voters$. If for each $c \in S$, $d(v, c) \ge t$, then we are done. Otherwise, suppose that there exists a candidate, wlog, say $c_i \in S$, such that $d(v, c_i) < t$.  Now consider any other $c_i \in S$ distinct from $c_1$, and suppose also that $d(v, c_i) < t$. Then, $d(c_1, c_i) \le d(c_1, v) + d(c_i, v) < t + t < 2t$, which is a contradiction. Therefore, it must hold that for each $c_i \in S \setminus \left\{c_1\right\}$, $d(v, c_i) \ge t$, which implies that $S$ is a solution.
\qed \end{proof}

First, note that we can assume $\lambda+1 \le k$ -- otherwise $\lambda = k$ case can be easily solved in polynomial-time using the argument mentioned in the preliminaries. Now, if a set $S \subseteq \candidates$ with $|S| \ge \lambda+1$ satisfying the conditions of Observation \ref{obs1} exists, then we can immediately augment it with arbitrary $k-(\lambda+1)$ candidates from $\candidates \setminus S$, yielding a solution of size $k$. Thus, henceforth, we may assume that any subset $S \subseteq \candidates$ consisting of candidates that are pairwise $2t$ distance away from each other, has size at most $\lambda$. 

Let us fix $N$ to be one such subset -- note that we can compute $N$ in polynomial time using a greedy algorithm. We have the following observation that follows from the maximality of $N$.

\begin{observation}\label{obs2}
    Any point $p\in \candidates$ must be in $\bigcup_{c \in N} B(c, 2t)$. In other words, each $p \in \candidates$ is inside a ball of radius $2t$ centered at one of the points in $N$. 
\end{observation}


This simple observation, combined with the following covering-packing property of the underlying Euclidean space will allow our algorithm to pick points from the vicinity of those chosen by an optimal algorithm. 

\hide{
Next, we proceed towards describing the actual algorithm.

\begin{enumerate}[label=(\roman*),wide=0pt]
	
	\item We find (if we can) a subset $N\sse \candidates$ whose pairwise distance is at least $2t$. 
	
\end{enumerate}
}

\begin{proposition}($\clubsuit$)\label{prop:r2-balls}
	In $\Real^d$, for any $0<r_{1}< r_{2}$, a ball of radius $r_{2}$  can be covered by $\alpha_d \cdot (r_2/r_1)^d$ balls of radius $r_{1}$. Here, $\alpha_d$ is a constant that depends only on the dimension $d$. 
\end{proposition}
\hide{
\begin{proof}[Proof of Proposition~\ref{prop:r2-balls}]
    First, it is a well-known fact  that the volume of a $d$-dimensional ball of radius $r$ is $c_d \cdot r^d$, for some constant $c_d$ that depends on the dimension $d$. Now, imagine overlaying a grid of sidelength $\frac{r_2}{\sqrt{d}}$. From volume arguments, the number of grid cells partially or completely intersecting the ball of radius $r_2$ is $\frac{c_d r_1^d}{r_2^d / (\sqrt{d})^d}$. Now, a ball of radius $r_2$ centered at the center of each grid cell covers the entire cell. Thus, $\alpha_d \cdot (r_1/r_2)^{d}$ balls are sufficient to cover the entire ball of radius $r_1$.
\qed \end{proof}
}
Next, for each $c\in N$, we find an ``$\epsilon t/4$-net'' inside the ball $B(c, 2t)$, i.e., a maximal subset $Q \subseteq B(c, 2t) \cap \candidates$, such that (i) for any distinct $c_1, c_2 \in Q$, $d(c_1, c_2) > \epsilon t/4$, and (ii) For each $c_1 \in \candidates \setminus Q$, there exists some $c_2 \in Q$, such that $d(c_1, c_2) \le \epsilon t/4$. Note that $Q$ can be computed using a greedy algorithm. Next, we iterate over each $c' \in Q$, and mark the $\lambda-1$ closest unmarked candidates to $c'$ that are not in $Q$ (if any). Let $R_c \coloneqq Q \cup M$, where $M$ denotes the set of marked candidates during the second phase. 

\begin{observation}($\clubsuit$) \label{obs:net}
    For each $c \in N$, $|R_c| \le \Oh_{d}(\lambda \cdot (1/\epsilon)^d)$, where $\Oh_d(\cdot)$ hides a constant that depends only on the dimension $d$. 
\end{observation} 

Let $S' = \bigcup_{c \in N} R_c$. Finally, let $S \coloneqq N \cup S'$, and note that $|S| \le \mathcal{O}_d(\lambda^2 \cdot (1/\epsilon)^{d})$, where $\mathcal{O}_d(\cdot)$ notation hides constants that depend only on $d$. Now we consider two cases.

\begin{itemize}
	\item If $|S| \leq k$, then we augment it with arbitrary $k-|S|$ candidates from $\candidates \setminus S$, and output the resulting set. 
		
	\item If $|S| > k$, then try all possible $k$-sized subsets of $S$ to see if it constitutes a solution. There can be at most $\binom{|S|}{k} < 2^{|S|} = 2^{\Oh_d(\lambda^2(1/\epsilon)^{d})}$ sets to check resulting in time 
	$f(d,\lambda, 1/\epsilon) \cdot (|\voters|+|\candidates|)^{\Oh(1)}$, where $f(\cdot)$ is some function.
	\end{itemize}

The next lemma completes the proof. We prove it by comparing $S$ to an optimal solution, and show that for every point in the latter there is a point in the vicinity that is present in $S$. 

\begin{lemma}If $|S| > k$, then there exists a subset $Q \subseteq S$ of size $k$ that constitutes a solution.
\end{lemma}

\begin{proof} Suppose that there is an optimal solution, denoted by $O$,  that contains $k$ points and for each point $v\in \voters$ there exist at least $\lambda$ points in $O$ (called {\it representatives}, $\C{R}(v)$) that are at least $t$ distance away from $v$. Let $\C{R} = \bigcup_{v \in \voters} \C{R}(v)$ denote the set of all representatives. 
	
	First, due to \Cref{obs2}, each $c \in \C{R}$ is inside some $B(c', 2t)$ for some $c' \in N$. Let $\tilde{c} \in Q$ be the closest (breaking ties arbitrarily) candidate to $c$ from $Q$.
 By construction, $d(\tilde{c}, c) \le t\epsilon/4$. Let $A(\tilde{c}) \subseteq \C{R}$ be the points for which $\tilde{c}$ is the closest point in $R_{c'}$ (breaking ties arbitrarily). 
 
{\bf Case 1: $|A(\tilde{c})| \le \lambda$.} In this case, we claim that for each $c_1 \in A(\tilde{c})$, we have added a unique $c_2$ to $R_{c'} \subseteq S'$ such that $d(c_1, c_2) \le \epsilon t$. First, if $A(\tilde{c}) \subseteq R_{c'}$, then the claim is trivially true (the required bijection is the identity mapping). Otherwise, there exists some $c_1 \in A(\tilde{c})$ such that $c_1 \not\in R_{c}$. In particular, this means that $c_1$ was not marked during the iteration of the marking phase corresponding to $\tilde{c}\in Q$. This means that at least $\lambda-1$ other candidates with distance at most $\epsilon t/4$ from $\tilde{c}$ were marked. For any of these marked candidates $c_2$, it holds that $d(c_1, c_2) \le d(c_1, \tilde{c}) + d(\tilde{c}, c_2) \le \epsilon t/2 \le \epsilon t$. Accounting for $\tilde{c}$, this implies that, for each $c_1 \in A(\tilde{c})$, there are at least $\lambda \ge |A(\tilde{c})|$ distinct candidates in $R_c$ within distance $\epsilon t$.
Let $Q(\tilde{c}) \subseteq S'$ denote an arbitrary such subset of size $\lambda$ in this case.
	
{\bf Case 2. $|A(\tilde{c})| > \lambda$.} In this case, let $A'(\tilde{c}) \subset A(\tilde{c})$ be an arbitrary subset of size $\lambda$. We claim that $A'(\tilde{c})$ is sufficient for any solution. In particular, consider a $v \in \voters$ and $c \in A(\tilde{c}) \setminus A'(\tilde{c})$ such that $c$ is a representative of $v$. We claim that for all $c' \in A'(\tilde{c})$, $d(\tilde{c}, c') \ge (1-\epsilon/2)t$, which follows from $d(c, c') \le \epsilon t/2$. Thus, the $\lambda$ points of $A'(\tilde{c})$ constitute an approximate set of representatives for $v$. Now, by using the argument from the previous paragraph w.r.t. $A'(\tilde{c})$, we can obtain a set $Q(\tilde{c})$ of size $\lambda$, such that for any voter $v \in V$ such that $\C{R}(v) \cap A(\tilde{c}) \neq \emptyset$, every point in $Q(\tilde{c})$ is at distance at least $(1-\epsilon)t$ from $v$. 
	
	Finally, let $Q$ denote the union of all sets $Q(\tilde{c})$ defined in this manner (note that $Q(\tilde{c})$ is defined only if $A(\tilde{c}) \neq \emptyset$). First, by construction, for each $v \in \voters$, $Q$ contains at least $\lambda$ points at distance at least $(1-\epsilon)t$. Next, $Q \subseteq R'$ and $|Q| \le k$ since for each point in $\C{R}$, we add at most one point to $Q$. Now, if $|Q| < k$, then we can simply add arbitrary $k-|Q|$ points to obtain the desired set. 
\hide{
}
	\qed \end{proof}


In fact, the covering-packing properties of the underlying metric space that are crucial in our algorithm are abstracted in the following well-known notion.

\begin{definition}({\bf Doubling dimension and doubling spaces})
	Let $\X{M} = (P, d)$ be a metric space, where $P$ is a set of points and $d$ is the distance function. We say that $\X{M}$ has \emph{doubling dimension} $\delta$, if for any $p \in P$, and any $r \ge 0$, the ball $B(p, r) \coloneqq \{q \in P: d(p, q) \le r\}$ can be covered using at most $2^\delta$ balls of radius $r/2$. If the doubling dimension of a metric space $\X{M}$ is a constant, then we say that $\X{M}$ is a \emph{doubling space}.
\end{definition}
Note that Euclidean space of dimension $d$ has doubling dimension $O(d)$. By a simple repeated application of the above definition, we get the following proposition that is an analogue of \Cref{prop:r2-balls}.
\begin{proposition} \label{prop:doubling}
	Let $\X{M} = (P, d)$ be a metric space of doubling dimension $\delta$. Then, any ball $B(p, r_2)$ can be covered with $\left(\lceil \frac{r_2}{r_1} \rceil\right)^{\delta}$ balls of radius $r_1$, where $0 < r_1 \le r_2$.
\end{proposition}

Our algorithm generalizes to metric spaces of doubling dimension $\delta$ in a straightforward manner, resulting in the following theorem. 

\begin{theorem}\label{thm:fpt-as}
For any $\epsilon$, $0< \epsilon < 1$, we have an algorithm that given
	an instance of \problemshort in a metric space of doubling dimension $\delta$, computes a solution of size $k$ such that for every point $v\in \voters$ there are at least $\lambda$ points in the solution that are at distance at least $(1-\epsilon)t$ from $v$ in time \fpt in $(\epsilon, \lambda, \delta)$. In particular, we obtain this result in Euclidean spaces of dimension $d$, in time \fpt in $(\epsilon, \lambda, d)$.
\end{theorem}

\section{Outlook}

In this paper we studied  a committee selection problem, where preferences of voters towards candidates was captured via a metric space.  In particular, we studied a variant where larger distance corresponds to higher preference for a candidate in comparison to a candidate who is farther away. We showed that our problem is NP-hard in general, and designed some polynomial time algorithms, as well as (parameterized) approximation algorithms. We conclude with some research directions for future study. One of our concrete open question is that Is \problemshort in  $\B{R}^{2}$ for $\lambda>1$ polynomial-time solvable? In this paper, we considered median scoring rules. It would be interesting to study other scoring rules as well when the voters and candidates are embedded in a metric space.
%
Moreover, we note that situations where we want an exact number of facilities to be built near each neighborhood and others to be far\hide{such as one where a high traffic generating facility,} is not handled by our model and would be worthy of future work. This would constitute the ``exact'' variant of our problem \problemshort and would be of natural interest. 


\newpage 
\bibliographystyle{splncs04}
\bibliography{comsoc_clustering}

\appendix

\section{Missing Proof of Section~\ref{sec:egal-cc}}

\begin{proof}[Proof of Lemma~\ref{lemma:dp-correctness}]
    The proof is by induction on $i$. First, for convenience we reiterate the properties of the region $R(x, y, p, c_1, c)$ corresponding to an entry $A[x, y, p, c, i$]:
	\begin{itemize}
		\item $R = R(x, y, p, c_1, c)$ is a convex region bounded by a set $\cA(R(x, y, p, c_1, c))$ of $i-1$ circular arcs, and straight-line segment $\seg{xy}$, such that $\cA(R)$ contains:
		\begin{itemize}
			\item At most one arc of the form $\arc(\cdot, \cdot, c')$ for every $c' \in \candidates$.
			\item Exactly one arc of the form $\arc(x, \cdot, c_1)$, which is the \emph{first} arc traversed along the boundary of $R(x, y, p, c_1, c, i)$ in clockwise direction, starting from $x$.
			\item $\arc(y, p, c)$ 
		\end{itemize}
		\item $R \cap \voters = \emptyset$.
	\end{itemize}
	
	For the base case, we consider all entries with $i \le 3$, and note that we fill the corresponding table entries by explicitly enumerating all candidate regions bounded by $i-1$ arcs that satisfy the corresponding properties. Thus, the correctness of such an entry is witnessed by the region we construct.
	
	Now, we inductively assume that the claim holds for some $i \ge 3$, and prove that it also holds for $i+1$. Fix such an entry $A[x, y, p, c_1, c, i+1]$, with $i+1 \ge 4$. 
	
	\textbf{Forward direction.} Suppose that $A[x, y, p, c_1, c, i+1] = \btrue$. Then, by construction, the following two conditions hold: 
	\begin{enumerate}
		\item $R(x, y, p, c_1, c) \cap \voters = \emptyset$, where, recall that $R(x, y, p, c_1, c)$ is the region bounded by $\seg{xp}, \seg{xy}$, and $\arc(y, p, c)$, and
		\item For some $z \in \cP$, and some $c' \in \candidates$ with $c' \not\in \{c, c_1\}$, it holds that
		\begin{itemize}
			\item $\arc(p, z, c')$ exists, 
			\item When traversing along this arc from $z$ to $p$, the $\arc(p, y, c)$ is a ``right turn'' (as formally defined in the algorithm description), and
			\item $A[x, p, z, c_1, c', i] = \btrue$.
		\end{itemize}
	\end{enumerate}
	
	By induction, since $A[x, p, z, c_1, c', i] = \btrue$, we know that there exists a region $R' = R(x, p, z, c_1, c', i)$ satisfying the required properties. Let $\cA' = \{\arc(x = u_0, u_1, c_1), \arc(u_1, u_2, c_2), \ldots, \arc(u_{i-2} = z, u_{i-1} = p, c_{i-1} = c')\}$ be the set of arcs bounding $R'$, along with the segment $\seg{xp}$. 
	
	First, we observe using inductive hypothesis, that $R'$ is contained in the region $R^*$ defined by the following halfplanes:
	\begin{itemize}
		\item For each $1 \le j \le i-2$, consider the tangents $\ell_{j, j}$, and $\ell_{j, j+1}$ at the point $u_{j}$ to the circles $D(c_{j})$ and $D(c_{j+1})$, respectively. Then, the halfplanes $H_{j, j}$ (resp. $H_{j, j+1}$) is the closed halfplane defined by $\ell_{j, j}$ (resp. $\ell_{j, j+1}$) that contains $c_{j}$ (resp. $c_{j+1}$).
		\item The halfplane $H_{0, 1}$ defined by tangent $\ell_{0, 1}$ at $x = u_0$ to $D(c_1)$ that contains $c_1$.
		\item The halfplane $H_{i-1, i-1}$ defined by the tangent $\ell_{i-1, i-1}$ at $p = u_{i-1}$ to $D(c_{i-1})$, that contains $c_{i-1}$.
		\item The halfplane $H_{xp}$ defined by line containing $\seg{xp}$ that contains (e.g.,) $z$. 
	\end{itemize}
	
	First, we show the following.
	\begin{clm}
		$\cA'$ does not contain an arc from the boundary of $D(c)$. 
	\end{clm}
	\begin{proof}
		Suppose otherwise, and that $\cA'$ contains some $a' = \arc(u_{j-1}, u_{j}, c_{j})$, for some $1 \le j \le i-1$, such that $c_{j+1} = c$. By construction, $j \neq i-1$, since $c' \neq c$. Similarly, due to aforementioned modification, $c' \neq c_1$. 
		
		
		Now, suppose that $1 < j < i-1$. However, the only part of the boundary of $D(c_j)$ that is contained in $R^*$ is the $\arc(u_{j-1}, u_j, c_j)$. This implies that the point $p$ lies on $\arc(u_{j-1}, u_j, c_j)$. However, this cannot happen if we traverse the boundary of $R'$ using arcs of $\cA'$ in the clockwise manner, and each turn is a right turn. This completes the proof of the claim.
	\qed \end{proof}
	
	Now, consider the set of arcs $\cA \coloneqq \cA' \cup \{\arc(y, p, c)\}$, and the region $R$ bounded by it. First, we observe that $R = R' \uplus R(x, y, p, c_1, c)$. Since $R' \cap \voters = \emptyset$ by induction, and $R(x, y, p, c_1, c) \cap \voters = \emptyset$ by construction, it holds that $R \cap \voters = \emptyset$. Therefore, $R$ satisfies all the conditions for the region $R(x, y, p, c_1, c, i)$, which completes the forward direction.
	
	\textbf{Reverse direction.} This proof is very much analogous to the forward direction. Suppose that there exists a region $R = R(x, y, p, c_1, c)$ satisfying the requirements, in particular that $R \cap \candidates = \emptyset$. Then, we consider $R = R' \uplus R(x, y, p, c_1, c)$, where $R'$ is the region bounded by the first $i-2$ arcs that are same as $R$, and the segment $\seg{xp}$. Let $\arc(p,z, c')$ be the last arc defining $R'$. Then, the entry $A[x, p, z, c_1, c', i-1]$ is \btrue using inductive hypothesis, and $c' \neq c$. Furthermore, $R(x, y, p, c) \cap \candidates = \emptyset$. Therefore, by construction we will set $A[x, y, p, c_1, c, i] = \btrue$. This completes the proof of Lemma~\ref{lemma:dp-correctness}.
\qed
\end{proof}

\begin{proof}[Proof of Lemma~\ref{lem:hittingset-equiv}]
	In the forward direction, let $H\subseteq \cU$ be a hitting set of size at most $k$. By adding arbitrary elements, we obtain $H' \subseteq \cU$ of size exactly $k$, which remains a hitting set. Let $H''$ be the subset of $\candidates$ corresponding to $H'$. Now, for any $S \in \cF$, $H$ contains an element $e$ such that $e \in S$. Thus, for every $v_S \in \voters$, there exists a $c_e \in H''$, such that $d(c_e, v_S) = 3$ by Observation \ref{obs:hittingset-equiv}. 
	
	In the reverse direction, let $H \subseteq \candidates$ be a set of size $k$ such that for any $v_S \in \voters$,  $\max_{c_e \in H} d(c_e, v_S) = 3$. Let $H'$ be the corresponding subset of elements. 
	For every $v_S \in \cF$, $H$ contains a candidate $c_e$ such that $d(c_e, v_S) = 3$. By Observation \ref{obs:hittingset-equiv}, $e \in S$. This implies that $H'$ is a hitting set.
\qed \end{proof}

\section{Missing proofs of Section~\ref{sec:large value}}

\begin{proof}[Proof of Proposition~\ref{prop:r2-balls}]
    First, it is a well-known fact  that the volume of a $d$-dimensional ball of radius $r$ is $c_d \cdot r^d$, for some constant $c_d$ that depends on the dimension $d$. Now, imagine overlaying a grid of sidelength $\frac{r_2}{\sqrt{d}}$. From volume arguments, the number of grid cells partially or completely intersecting the ball of radius $r_2$ is $\frac{c_d r_1^d}{r_2^d / (\sqrt{d})^d}$. Now, a ball of radius $r_2$ centered at the center of each grid cell covers the entire cell. Thus, $\alpha_d \cdot (r_1/r_2)^{d}$ balls are sufficient to cover the entire ball of radius $r_1$.
\qed \end{proof}

\begin{proof}[Proof of Observation~\ref{obs:net}]
    We prove that the size of the set $Q$ is bounded by $\Oh_{d}((1/\epsilon)^d)$. Since for each $c' \in Q$ we mark at most $\lambda -1$ additional candidates, the bound on $|R_c|$ follows.
    
    From Proposition \ref{prop:r2-balls}, the ball $B(c, 2t)$ can be covered using at most $\alpha_d \cdot (16/\epsilon)^d$ balls of radius $\epsilon t/8$. Let $Z$ denote the centers of these balls. We construct a mapping $\varphi: Z \to Q \cup \left\{\bot\right\}$ for showing the bound on $|Q|$.  
    
    Fix some $p \in Z$. First we argue that there exists at most one point $c_1 \in Q$ such that $d(p, c_1) \le \epsilon t/8$. Suppose not. Then exist two distinct $c_1, c_2 \in Q$ such that $d(p, c_1), d(p, c_2) \le \epsilon t/8$, then this implies $d(c_1, c_2) \le \epsilon t/4$, which is a contradiction. Now, if for $p \in Z$ there exists a $c_1 \in Q$ such that $d(p, c_1) \le \epsilon t/8$, then define $\varphi(p) = c_1$. Otherwise, if all $c' \in Q$ are at a distance more than $\epsilon t/8$ away from $p$, then define $\varphi(p) = \bot$. 
    
    Finally, note that $Q \subseteq B(c, 2t)$, and the balls of radius $\epsilon t/8$ around the points of $Z$ cover the entire ball $B(c, 2t)$. Therefore, for each $c \in Q$, there exists at least one $p \in Z$ such that $\varphi(p) = c$.  Hence, $|Q| \le |Z|$, which completes the proof.
\qed \end{proof}

\end{document}